\documentclass[preprint,12pt]{elsarticle}
\usepackage{graphics, epsfig, comment, color}

\def\be{\begin{equation}}
\def\ee{\end{equation}}
\def\bea{\begin{eqnarray}}
\def\eea{\end{eqnarray}}
\def\l{\left}
\def\r{\right}

\def\h{\frac{1}{2}}
\def\det{\sqrt{-g}}

\newcommand{\nn}{\nonumber}

\begin{document}

\title{The dynamics of metric-affine gravity}
\author[vitlib]{Vincenzo Vitagliano} \ead{vitaglia@sissa.it}
\author[sot]{Thomas P. Sotiriou} \ead{T.Sotiriou@damtp.cam.ac.uk}
\author[vitlib]{Stefano Liberati} \ead{liberati@sissa.it}
\address[vitlib]{SISSA-International School for Advanced Studies, Via Bonomea
265, 34136 Trieste, Italy and  INFN, Sez. di
Trieste, Via Valerio 2, 34127 Trieste, Italy}
\address[sot]{Department of Applied Mathematics and Theoretical Physics, Centre for Mathematical Sciences, University of Cambridge, Wilberforce Road, Cambridge, CB3 0WA, UK}

\begin{abstract}
Metric-affine theories of gravity provide an interesting alternative to General Relativity: in such an approach, the metric and the affine (not necessarily symmetric) connection are independent quantities. Furthermore, the action should include covariant derivatives of the matter fields, with the covariant derivative naturally defined using the independent connection. As a result, in metric-affine theories a direct coupling involving matter and connection is also present. The role and the dynamics of the connection in such theories is explored. We employ power counting in order to construct the action and search for the minimal requirements it should satisfy for the connection to be dynamical. We find that for the most general action containing lower order invariants of the curvature and the torsion the independent connection does not carry any dynamics. It actually reduces to the role of an auxiliary field and can be completely eliminated algebraically in favour of the metric and the matter field, 
introducing extra interactions with respect to general relativity. However, we also show that including higher order terms in the action radically changes this picture and excites new degrees of freedom in the connection, making it (or parts of it) dynamical. Constructing actions that constitute exceptions to this rule requires significant fine tuned and/or extra {\em a priori} constraints on the connection. We also consider $f(\mathcal{R})$ actions as a particular example in order to show that they constitute a distinct class of metric-affine theories with special properties, and as such they cannot be used as representative toy theories to study the properties of metric-affine gravity. 
\end{abstract}

\maketitle

\section{Introduction}

In general relativity spacetime geometry is fully described by the metric. That is to say, the metric does not only define distances, which is its primary role, but also defines parallel transport, as it is used to construct the Levi-Civita connection. However, in principle this does not have to be the case. The metric and the connection can be independent quantities. In this case one would need field equations that would determine the dynamics of both the metric and the connection.

How can one construct such a theory?
In some textbooks, see for example Refs.~\cite{grav,wald}, an independent variation with respect to the metric and the connection of what is formally the Einstein--Hilbert action is considered as an alternative way to arrive to Einstein's equations. This variation is called Palatini variation. Indeed the variation with respect to the connection leads to a non-dynamical equation fixing the latter to be equal to the Levi-Civita connection of the metric, and under this condition the field equations for the metric become Einstein's equations. However, there is a very crucial implicit assumption: that the matter action does not depend on the connection. This is equivalent to assuming that covariant derivatives of matter fields are defined with the Levi-Civita connection of the metric instead of the independent connection. Then the independent connection does not really carry the geometrical meaning described previously, see Refs.~\cite{sot2,T&S} for a discussion. 

Note that if one allows the connection to enter the matter action, the resulting theory is no longer general relativity \cite{hehlkerl}.
Additionally, one can easily argue that within a metric-affine setting the Einstein--Hilbert form of the action is not necessarily well motivated anyway: under the assumption that the connection is the Christoffel symbol of the metric, the Einstein--Hilbert action is indeed the unique diffeomorphism invariant action which leads to second order field equations (modulo topological terms and total divergences). However, this is not the case if the connection is allowed to be independent and it is not assumed to be symmetric: in this case there are other invariants one should in principle include in the action, even with the same dimensions as the Ricci scalar.

The situation gets more complicated once one decides to consider the role of higher order terms. Again, such actions have been studied mostly under the simplifying but geometrically unappealing assumption that the connection does not enter the matter action. Such theories are dubbed Palatini theories of gravity. Particular attention has been payed to $f({\cal R})$ models, i.e.~to actions where the Lagrangian is some algebraic function of the Ricci scalar of the independent connection, ${\cal R}$. Such actions were introduced and initially studied by Buchdahl \cite{buchdahl}. They have recently attracted a lot of interest as possible infrared modifications of general relativity \cite{palgen} (see Refs.~\cite{T&V,deF,phd} for reviews). However, Palatini $f({\cal R})$ gravity models with infrared corrections with respect to GR have been shown to be non-viable for various reasons: they are in conflict with the standard model of particle physics \cite{flanagan,padilla} and they violate solar system tests as their 
post-Newtonian metric  has an algebraic dependence on the matter fields \cite{olmonewt,sotnewt}. Singularities have been shown to arise on the surface of well known spherically symmetric matter configurations \cite{Barausse:2007pn}, which render the theory at best incomplete and provide a very strong viability criterion. This criterion is almost independent of the functional form of the Lagrangian, the only exception being Lagrangians with corrections which become important only in the far ultraviolet (as in this case the singularities manifest at scales where non-classical effects take over) \cite{Olmo:2008pv} .
Generalized Palatini theories of gravity have also been considered. For example, Lagrangians of the form $f({\cal R}^{(\mu\nu)}{\cal R}_{(\mu\nu)})$ (parentheses indicating symmetrization) were studied in Ref.~\cite{Allemandi:2004wn} and Lagrangians of the form ${\cal R}+f({\cal R}^{(\mu\nu)}{\cal R}_{(\mu\nu)})$ were considered in Ref.~\cite{Li:2007xw}, with attention being focussed on cosmology. In Refs.~\cite{Olmo:2009xy,olmo2,new} Lagrangians of the more general form $f(R,R^{\mu\nu}R_{\mu\nu})$ were studied, see however Ref.~\cite{STV}.

Unlike the exceptional case of the Einstein--Hilbert action as mentioned above, generalized Palatini theories of gravity are distinct from the theories one would get starting from the same action (formally) and applying standard metric variation. One cannot say that their dynamics have been well understood in general. That is because the dynamic of the most well studied class, $f({\cal R})$ are rather peculiar and not representative. Indeed, in Palatini $f({\cal R})$ gravity the independent connection does not carry any dynamics and can be algebraically eliminated in favour of the metric and the matter fields \cite{sotplb,Sotiriou:2007zu}. This result has recently been generalized to  $f(R)$ theories with non-symmetric connections, {\em i.e.}~theories that allow for torsion \cite{Sotiriou:2009xt}. The lack of extra dynamics with respect to general relativity can also be seen by the fact that Palatini $f(R)$ gravity has been shown to be dynamically equivalent to Brans--Dicke theory with Brans--Dicke parameter 
$\omega_0=-3/2$ \cite{flanagan,olmonewt,sot1} (again, irrespectively of how general the connection is allowed to be \cite{Sotiriou:2009xt}). This is a particular theory within the Brans--Dicke class in which the scalar does not carry any dynamics and can be algebraically eliminated in favour of the matter fields. 

The algebraic elimination of the connection (or the corresponding scalar field in the Brans--Dicke representation) will introduce extra matter interactions \cite{flanagan,padilla} and Palatini $f({\cal R})$ theories will essentially be equivalent to general relativity with modified source terms. In fact, this property is what lies in the heart of all the viability issues mentioned earlier \cite{Barausse:2007pn}. However, this is not a generic property of generalized Palatini gravity, as it has been recently demonstrated in Ref.~\cite{STV}, but just a peculiarity of $f({\cal R})$ actions. Generic higher order actions lead to extra dynamical degrees of freedom.

What we would like to understand here is what happens when one jumps from the Palatini approach, to the more general and better motivated metric-affine approach, where the independent connection is allowed to enter the matter action, define the covariant derivative, and, therefore, retain its geometrical significance. In particular, we would like to understand under which circumstances this connection becomes an auxiliary field, which can be algebraically eliminated, and when it actually does carry dynamics. Note that there are well known examples, such as Einstein--Cartan theory \cite{hehl} (which is a metric-affine theory with the additional constraint that the connection is metric, but not symmetric), where the independent connection can be eliminated algebraically, leading to general relativity with extra matter interactions. In this specific case, this is a four-fermion interaction. See also Ref.~\cite{yuri} for an example of a more general action with the same property. What happens for more general 
theories, however, and especially how the dynamics of the connection will be affected by considering higher order terms in the action, has not been systematically understood.

In order to address this issue we follow an approach motivated by effective field theory. We will consider the metric-affine action as an effective action, possibly arising from some more fundamental theory at some appropriate limit. We will then employ power counting in order to construct the most general action order by order. This will allow us to arrive at  model independent statements and avoid considering fine-tuned actions, which can lead to misleading results. 

The paper is organized in the following way. Section \ref{general} is devoted to presenting our conventions and briefly reviewing the metric-affine set up. In section \ref{2order} we construct the most general second order action and show that it does not lead to a dynamical connection. In section \ref{higher} we move on to consider actions with higher order invariants and we show that, remarkably, the situation changes radically and degrees of freedom residing in the connection become excited. In section \ref{fofr} we consider $f({\cal R})$ actions as a particular example, as they have been extensively studied in the literature, even in the metric-affine setting \cite{T&S}. We do not consider them because we expect them to be representative examples of the dynamics of metric-affine gravity. On the contrary, our intention is to explicitly demonstrate that they are not. Section \ref{discuss} contains a discussion of our results and our conclusions.


\section{General setup for metric-affine theories}
\label{general}
We start by clarifying our notation and conventions. The covariant derivative of the connection $\Gamma_{\;\;\mu\nu}^{\rho}$ acting on a tensor is defined as
\bea\label{def}
\nabla_\mu A^\nu_{\phantom{a}\sigma}=\partial_\mu A^\nu_{\phantom{a}\sigma}+\Gamma^\nu_{\phantom{a}\alpha\mu} A^\alpha_{\phantom{a}\sigma}-\Gamma^\alpha_{\phantom{a}\sigma\mu} A^\nu_{\phantom{a}\alpha}\,.
\eea
It is important to stress that the position of indices must be taken very carefully into account, since the connection are not assumed to be symmetric. The antisymmetric part of the connection is commonly referred to as the Cartan torsion tensor 
\bea
\label{cartan}
S_{\mu\nu}^{\phantom{ab}\lambda}\equiv \Gamma^{\lambda}_{\phantom{a}[\mu\nu]}\,.
\eea
The failure of the connection to covariantly conserve the metric is measured by the non-metricity tensor
\bea\label{nonmet}
Q_{\lambda\mu\nu}\equiv-\nabla_{\lambda}g_{\mu\nu}\,.
\eea

Using the  connection one can construct the Riemann tensor
\be
\label{riemann}
{\cal R}^\mu_{\phantom{a}\nu\sigma\lambda}=-\partial_\lambda\Gamma^\mu_{\phantom{a}\nu\sigma}+\partial_\sigma\Gamma^\mu_{\phantom{a}\nu\lambda}+\Gamma^\mu_{\phantom{a}\alpha\sigma}\Gamma^\alpha_{\phantom{a}\nu\lambda}-\Gamma^\mu_{\phantom{a}\alpha\lambda}\Gamma^\alpha_{\phantom{a}\nu\sigma}\, .
\ee
which has no dependence on the metric. Notice that the Riemann tensor here has only one obvious symmetry: it is antisymmetric in the last two indices. All other symmetries one might by accustomed to from general relativity are not present for an arbitrary connection \cite{schro}.
Without any use of the metric we can also define as $\mathcal{R}_{\mu\nu}$ the Ricci tensor built with the connection $\Gamma_{\;\;\mu\nu}^{\rho}$
\bea\label{ricci}
{\cal R}_{\mu\nu}\equiv {\cal R}^\lambda_{\phantom{a}\mu\lambda\nu}=\partial_{\lambda}\Gamma^{\lambda}_{\;\;\mu\nu}-\partial_{\nu}\Gamma^{\lambda}_{\;\;\mu\lambda}+\Gamma^{\lambda}_{\;\;\sigma\lambda}\Gamma^{\sigma}_{\;\;\mu\nu}-\Gamma^{\lambda}_{\;\;\sigma\nu}\Gamma^{\sigma}_{\;\;\mu\lambda}\,.
\eea
$\mathcal{R}=g^{\mu\nu} {\cal R}_{\mu\nu}$ is the corresponding Ricci scalar. 

Note that there is an intrinsic ambiguity in the definition of the Ricci tensor in metric-affine theories as the limited symmetries of the Riemann tensor allow now for an alternative definition as 
\be
\hat{{\cal R}}_{\mu\nu}\equiv {\cal R}^\sigma_{\phantom{a}\sigma\mu\nu}=\,-\partial_\nu\Gamma^\sigma_{\phantom{a}\sigma\mu}+\partial_\mu\Gamma^\sigma_{\phantom{a}\sigma\nu}\,.
\ee
This tensor is called the homothetic curvature.
For a symmetric connection it is equal to the antisymmetric part of ${\cal R}_{\mu\nu}$ and, therefore, it need not be separately considered. This is not the case for a non-symmetric connection. Note however, that the homothetic curvature is fully antisymmetric and as such it leads to a vanishing scalar when contracted with the metric.\footnote{See Ref.~\cite{T&S} for a more detailed discussion about the ambiguities in the definition of the Ricci tensor. Note also that, unlike the usual Ricci tensor, the homothetic curvature tensor has a direct physical interpretation: it measures the change of the length of a vector when it is transported along a closed loop. When the homothetic curvature vanishes, the connection is volume
preserving, {\em i.e.}~lengths and volumes do not change during parallel transport. We thank Yuri Obukhov for bringing this to our attention.}

As already mentioned in the Introduction, the key characteristic of metric-affine gravity is that the affine connection $\Gamma_{\;\;\mu\nu}^{\rho}$ is not assumed to have any {\em a priori} relation with the metric. On the other hand, it is assumed to define parallel transport and the covariant derivative of matter fields, so it inevitably enters the matter action, see Ref.~\cite{sot2} for a discussion. That is, in metric-affine gravity couplings between the connection and the matter fields are allowed. This is the main difference from (generalized) Palatini theories of gravity, as mentioned earlier. The action will, therefore be of the following general form
\begin{equation}
 S=S_G+S_M=\int d^4x\det \left[ \mathcal{L}_G(g_{\mu\nu}, \Gamma_{\;\;\mu\nu}^{\rho})+\mathcal{L}_M\l(g_{\mu\nu}, \Gamma_{\;\;\mu\nu}^{\rho}, \psi\r)\right]\,,
\end{equation}
where $g$ is the determinant of the metric $g_{\mu\nu}$, $\psi$ collectively denotes the matter fields, and $S_M$ is the matter action. We have written the dependence of ${\cal L}_M$ on the various fields explicitly to avoid confusion here, but we will suppress it from now on and just use $S_M$ instead in order to lighten the notation. Clearly, specific choices of matter fields will not couple to the connection of course, such as a scalar field or a gauge field. See Ref.~\cite{T&S} for a detailed discussion on such matters and on the general characteristics of metric-affine gravity.

One now needs to specify the exact form of the Lagrangian $\mathcal{L}_G$. In Ref.~\cite{hehlkerl} an action linear in ${\cal R}$ was consider and in Ref.~\cite{T&S} the most general $f({\cal R})$ family was studied extensively. Instead of an {\em ad hoc} choice inspired by some similarity with the Einstein--Hilbert action and its generalizations, we would like to follow here an effective field theory approach in order consider the most general action possible at each order. To construct this action, we should carry on a power counting analysis which will reveal the whole set of appropriate terms order by order. We set $c=1$ and we can choose the engineering dimensions 
\be
[dx]=[dt]=[l]
\ee
where $l$ is a place holder symbols with dimension of a length. Then we have
\bea
&&[g_{\mu\nu}]=[1]\,, \quad [\sqrt{-g} dx^4]=[l^4]\,, \quad[\Gamma^{\lambda}_{\phantom{a}\mu\nu}]=[l^{-1}]\,,\quad [{\cal R}_{\mu\nu}]=[l^{-2}]\,.
\eea
Now consider as a simple example the action
\be
\label{action}
S_G=\frac{1}{l_p^2} \int dx^3 dt  \sqrt{-g} {\cal R}\,.
\ee
Requiring that this action is dimensionless implies that the coupling constant $l_p$ must have dimensions of a length and can be naturally associated to the Planck length. What we mean by order of the gravitational theory is also clear now:  we mean the highest order in $l_p^{-1}$ powers appearing in the Lagrangian (which, since one cannot choose the metric and the connection to be dimensionless at the same time, does not correspond to the order in its derivatives).


\section{Second order action}
\label{2order}

Clearly the action written above is not the most general one we could write in metric-affine gravity. It is just an example inspired by the analogy with standard GR and the Einstein--Hilbert action. To begin with, we could include a cosmological constant term, which is of lower order. But such a term would not play any important role in our arguments so we will omit it for simplicity. What other terms can we write at the second order? Under the assumption that the connection is torsionless and metric compatible (Levi-Civita), there exist no other term which respects diffeomorphism invariance, as it is well known. But, in the more general metric-affine setting we are considering here, there is at least two more tensors one could imagine using in order to construct invariants: 
\begin{itemize}
\item The aforementioned ``second Ricci'' tensor $\hat{{\cal R}}_{\mu\nu}$. However, this tensor has dimensions $[l^{-2}]$ and is antisymmetric, so there is no quantity one can construct out of it at second order;
\item the Cartan torsion tensor of eq.~(\ref{cartan}), which has the same dimensions as $\Gamma^{\lambda}_{\phantom{a}\mu\nu}$. Therefore, terms with one derivative of $S_{\mu\nu}^{\phantom{ab}\lambda}$ or terms quadratic in $S_{\mu\nu}^{\phantom{ab}\lambda}$ will be of the same order as ${\cal R}$. 
\end{itemize}
 
 Due to the symmetries of $S_{\mu\nu}^{\phantom{ab}\lambda}$ there is only a single term with a derivative we can write
\be
\label{derterm}
g^{\mu\nu}\nabla_\mu S_{\nu\sigma}^{\phantom{ab}\sigma}\,.
\ee
For the same reason, there are just three terms quadratic in $S_{\mu\nu}^{\phantom{ab}\lambda}$ one can write
\be
\label{3terms}
g^{\mu\nu}S_{\mu\lambda}^{\phantom{ab}\lambda}S_{\nu\sigma}^{\phantom{ab}\sigma}\,, \qquad  g^{\mu\nu}S_{\mu\lambda}^{\phantom{ab}\sigma}S_{\nu\sigma}^{\phantom{ab}\lambda}\,, \qquad  g^{\mu\alpha}g^{\nu\beta}g_{\lambda\gamma}S_{\mu\nu}^{\phantom{ab}\lambda}S_{\alpha\beta}^{\phantom{ab}\gamma}\,.
\ee
Note that the term in eq.~(\ref{derterm}) has been considered by Papapetrou and Stachel in \cite{papa}.
  
A subtle point is the following.  The term in eq.~(\ref{derterm}) is not a total divergence as $\nabla_\mu$ is not defined with the Levi-Civita connection of the metric. On the other hand, one can think to decompose the connection as
\be\label{decomp}
\Gamma^{\lambda}_{\phantom{a}\mu\nu}= \left\{^{\lambda}_{\phantom{a}\mu\nu}\right\}+C^{\lambda}_{\phantom{a}\mu\nu}\,,
\ee
i.e.~in its Levi-Civita part and the rest. Now, using this decomposition we can split the covariant derivative in (\ref{derterm}) in a metric compatible part, which will lead to a total divergence, and the rest, which will lead to terms consisting of contractions between $C^{\lambda}_{\phantom{a}\mu\nu}$ and the Cartan torsion tensor. Since the non-metricity is not zero, these terms are different than the ones already considered above in eq.~(\ref{3terms}). Thus the term in eq.~(\ref{derterm}) is non-trivial.
 
This brings us to another puzzle though: $C^{\lambda}_{\phantom{a}\mu\nu}$ is a tensor, so why not consider terms constructed with it as well? Actually, $C^{\lambda}_{\phantom{a}\mu\nu}$ can always be decomposed in terms of torsion $S_{\mu\nu}^{\phantom{ab}\lambda}$ and non-metricity $Q_{\lambda\mu\nu}$, so the question then reduces to whether we should also consider terms constructed with $Q_{\lambda\mu\nu}$ or not. From a power counting/field theory perspective nothing prevents us from doing so, and these would indeed be terms of the same order. However, from this perspective we should also consider, for instance, the Ricci tensor of the metric $R_{\mu\nu}$. In fact, $Q_{\lambda\mu\nu}$ and $R_{\mu\nu}$ share a common characteristic which is crucial for our discussion: They cannot be expressed without using derivatives of the metric (even if instead of (\ref{nonmet}) one tries to define $Q_{\lambda\mu\nu}$ using the connection, then still the Levi-Civita connection would be needed as well). Therefore, the 
puzzle reduces to whether  or not we should be considering invariants constructed with derivatives of the metric. 
 
Clearly, field theoretic considerations cannot give an answer to this question. Such terms should be considered unless we are willing to invoke some principle excluding them, along the line of minimal coupling in general relativity. Such a principle has been discussed in Ref.~\cite{T&S}. In simple terms it would be the requirement that the metric be used only for raising and lowering indices. We choose to follow this prescription here, as it seems sensible from a geometrical perspective (the purpose of the metric being to measure distances) and it significantly reduces the number of terms one can consider. 

Another way to reduce the number of terms without invoking a minimal coupling principle would be to require that the connection be metric compatible. This would force $Q_{\lambda\mu\nu}$ to vanish, without necessarily implying torsion has to vanish as well. We would then remain with exactly the same terms written above. However, in this case the term in eq.~(\ref{derterm}) would indeed differ from the first term in eq.~(\ref{3terms}) only by a total surface term and one would be able to omit it.


Let us consider the consider the most general second-order action we have just constructed in our setting
\bea\label{2ndorder}
S=\frac{1}{16\,\pi\,l_p^2} \int dx^4  \sqrt{-g} &&\l(g^{\mu\nu}{\mathcal R}_{\mu\nu}+a_1g^{\mu\nu}\nabla_{\mu}S_{\nu\sigma}^{\;\;\;\;\sigma}+a_2g^{\mu\nu}S_{\mu\lambda}^{\;\;\;\;\lambda}S_{\nu\sigma}^{\;\;\;\;\sigma}\r.\\
&& \l.+a_3g^{\mu\nu}S_{\mu\lambda}^{\;\;\;\;\sigma}S_{\nu\sigma}^{\;\;\;\;\lambda}+a_4g^{\mu\alpha}g^{\nu\beta}g_{\lambda\gamma}S_{\mu\nu}^{\;\;\;\;\lambda}S_{\alpha\beta}^{\;\;\;\;\gamma}\r)+S_M\,,\nn
\eea
where the $a_i$'s represent the various coupling constant. Varying independently with respect to metric and connection yields
\bea
\label{maeq1}
&&{\cal R}_{(\mu\nu)}-\h g_{\mu\nu} \mathcal{R}+a_1\l\{\nabla_{(\mu}S_{\nu)}-\h g_{\mu\nu}g^{\alpha\beta}\nabla_{\alpha}S_{\beta}\r\}\nn\\
&&\quad +a_2\l\{-\h g_{\mu\nu} S_{\alpha}S^{\alpha}+S_{\mu}S_{\nu}\r\}+a_3\l\{-\h g_{\mu\nu} g^{\alpha\beta}S_{\alpha\lambda}^{\;\;\;\;\sigma}S_{\beta\sigma}^{\;\;\;\;\lambda}+S_{\mu\lambda}^{\;\;\;\;\sigma}S_{\nu\sigma}^{\;\;\;\;\lambda}\r\}\nn\\
&&\quad +a_4\l\{-\h g_{\mu\nu}S_{\rho\sigma\lambda} S^{\rho\sigma\lambda}+2 S_{\alpha\mu}^{\;\;\;\;\lambda}S^{\alpha}_{\phantom{a}\nu\lambda}-S_{\rho\sigma\mu}S^{\rho\sigma}_{\phantom{ab}\nu}\r\}=\kappa T_{\mu\nu}\,,\\
\label{vargamdef}
&&\frac{1}{\det}\l[-\nabla_\lambda\left(\sqrt{-g}g^{\mu\nu}\right)+\nabla_\sigma\left(\sqrt{-g}g^{\sigma\mu}\right)\delta^{\nu}_{\;\;\;\lambda}-a_1\nabla_{\alpha}\l(\det g^{\alpha[\mu}\r)\delta^{\nu]}_{\;\;\lambda}\r]\nn\\&&\quad+\l(2-a_1\r)g^{\mu\nu}S_{\lambda}-2S^{(\mu}\delta^{\nu)}_{\;\;\;\lambda}+2(a_1+a_2-1)S^{[\mu}\delta^{\nu]}_{\;\;\;\lambda}+2a_3g^{\alpha[\mu}S_{\alpha\lambda}^{\;\;\;\;\nu]}\nn\\
&&\quad +2a_4g^{\alpha[\mu}g^{\nu]\beta}g_{\lambda\gamma}S_{\alpha\beta}^{\;\;\;\;\gamma}{+2g^{\mu\sigma}S_{\sigma\lambda}^{\;\;\;\;\nu}}=\kappa \Delta_{\lambda}^{\;\;\mu\nu}\,.
\eea
where $S_{\alpha}\equiv S_{\alpha\beta}^{\;\;\;\;\beta}$, $\kappa=8\,\pi\,l_p^2$ and 
\be
T_{\mu\nu}\equiv-\frac{2}{\det}\frac{\delta S_M}{\delta g^{\mu\nu}}\,,\hspace{1.5cm}\Delta_{\lambda}^{\;\;\mu\nu}\equiv-\frac{2}{\det}\frac{\delta S_M}{\delta\Gamma_{\;\;\mu\nu}^{\lambda}}\,.
\ee
$\Delta_{\lambda}^{\;\;\mu\nu}$ is now as the \textit{hypermomentum} and, as already identified in \cite{hehl2}, it encapsulates all the information related to the spin angular momentum of matter, the intrinsic part of dilation current and the shear current. $T_{\mu\nu}$ on the other hand is sometimes referred to as the stress-energy tensor, in analogy with general relativity. However, it should be stressed that this terminology might be misleading within the metric-affine framework as this tensor does not have the properties usually associated with the stress-energy tensor is general relativity. For instance, it is not necessarily divergence free, it does not reduce to the special relativistic stress energy tensor at a suitable limit and of course it does  describe but some properties of matter, given the existence of $\Delta_{\lambda}^{\;\;\mu\nu}$ as well. In fact, it is best if it is just considered as nothing more than a short hand notation for the functional derivative of the matter action with respect 
to the metric.

Our present aim is to check whether it is possible to fully eliminate the connection from the field equations. Let us consider the contraction of $\lambda$ index in (\ref{vargamdef}) respectively with $\mu$ and $\nu$
\bea\label{nablag}
\frac{3}{2}\frac{a_1}{\det}\nabla_\mu\left(\sqrt{-g}g^{\mu\nu}\right)=\kappa\Delta_{\mu}^{\;\;\mu\nu}+(4a_1+3a_2+a_3+2a_4)S^{\nu}\,,
\eea
\bea
\frac{6-3a_1}{2 \det}\nabla_\nu\left(\sqrt{-g}g^{\mu\nu}\right)=\kappa\Delta_{\nu}^{\;\;\mu\nu}-(2a_1+3a_2+a_3+2a_4{-4})S^{\mu}\,.
\eea
Combining these two equations gives $S_\nu$ and the trace $\nabla_\mu\left(\sqrt{-g}g^{\mu\nu}\right)$ as functions of the hypermomentum
\bea
S^{\nu}=\frac{\kappa}{({2}-a_1)a_1+3a_2+a_3+2a_4} \l[(a_1-1)\Delta^{\nu}-\tilde{\Delta}^{\nu}\r]\,,
\eea
\bea\label{torsvec}
\frac{1}{\det}\nabla_\mu\left(\sqrt{-g}g^{\mu\nu}\right)&=&{\frac{2}{3}\l[\kappa \Delta^{\nu}+(a_1+2)S^{\nu}\r]=}\nn\\
&=&\frac{2}{3}\l\{\kappa\Delta^{\nu}+\frac{\kappa(a_1{+2})\l[(a_1-1)\Delta^{\nu}-\tilde{\Delta}^{\nu}\r]}{({2}-a_1)a_1+3a_2+a_3+2a_4} \r\}\,,
\eea
where we defined the two quantities $\Delta^{\mu}\equiv\Delta_{\alpha}^{\;\;(\alpha\mu)}$ and $\tilde{\Delta}^{\mu}\equiv\Delta_{\alpha}^{\;\;[\alpha\mu]}$. Eq. (\ref{torsvec}) can be inserted in (\ref{vargamdef}) to eliminate the second and the third term in order to get
\bea\label{newf}
&&\frac{1}{\det}\l[-\nabla_\lambda\left(\sqrt{-g}g^{\mu\nu}\right)\r]+2a_3g^{\alpha[\mu}S_{\alpha\lambda}^{\;\;\;\;\nu]}+2a_4g^{\alpha[\mu}g^{\nu]\beta}g_{\lambda\gamma}S_{\alpha\beta}^{\;\;\;\;\gamma}{+2g^{\mu\sigma}S_{\sigma\lambda}^{\;\;\;\;\nu}}=\nn\\
&&=\kappa \Delta_{\lambda}^{\;\;\mu\nu}\!\!-\frac{2}{3}\l[\kappa \Delta^{\mu}+(a_1{+2})S^{\mu}\r]\delta^{\nu}_{\;\;\;\lambda}\!+\frac{2}{3}a_1\l\{\kappa \Delta^{[\mu}+(a_1{+2})S^{[\mu}\r\}\delta^{\nu]}_{\;\;\;\lambda}-\nn\\&&\qquad-\l(2-a_1\r)g^{\mu\nu}S_{\lambda}+2S^{(\mu}\delta^{\nu)}_{\;\;\;\lambda}-2(a_1+a_2-1)S^{[\mu}\delta^{\nu]}_{\;\;\;\lambda}\,,
\eea
while we will refrain from replacing $S_\nu$ for compactness.

Using the identities
\be
\nabla_\mu{\sqrt{-g}}=\partial_\mu{\sqrt{-g}}-\Gamma^{\alpha}_{\;\;\alpha\mu}\sqrt{-g}
\ee
and
\begin{equation}
g_{\mu\nu}\partial_{\lambda}(\sqrt{-g}g^{\mu\nu})=2\sqrt{-g}\,\partial_{\lambda}\ln\sqrt{-g}
\end{equation}
we can write the trace of eq.~(\ref{newf}) with the metric in the $\mu$ and $\nu$ indices as 
\bea
\frac{\partial_{\lambda}\det}{\det}=-\h\kappa g_{\mu\nu}\Delta_{\lambda}^{\;\;\mu\nu}+\frac{1}{3}\kappa g_{\mu\lambda}\Delta^{\mu}+\l({\frac{8}{3}}-\frac{5}{3}a_1\r)S_{\lambda}+\Gamma^{\alpha}_{\;\;\alpha\lambda}\,.
\eea
Eliminating the density related terms and suitably lowering the indices eq.~(\ref{newf}) can eventually take the form
\bea
&&\partial_{\lambda}g_{\sigma\rho}-\Gamma^{\mu}_{\;\;\rho\lambda}g_{\mu\sigma}-\Gamma^{\nu}_{\;\;\sigma\lambda}g_{\nu\rho}+2a_3S_{[\sigma|\lambda|\rho]}+2a_4S_{\sigma\rho\lambda}{+2S_{\sigma\lambda\rho}}=\nn\\ 
&&=g_{\sigma\rho}\l({\frac{8}{3}}-\frac{5}{3}a_1\r)S_{\lambda}+\frac{1}{3}\kappa g_{\sigma\rho}\Delta_{\lambda}-\frac{1}{2}\kappa g_{\sigma\rho}\Delta_{\lambda\;\;\mu}^{\;\;\mu}+\nn\\&&\quad+\kappa\Delta_{\lambda\sigma\rho}-\l(2-a_1\r)g_{\sigma\rho}S_{\lambda}+2S_{(\sigma}g_{\rho)\lambda}-2(a_1+a_2-1)S_{[\sigma}g_{\rho]\lambda}+\nn\\ 
&&\quad+\frac{2}{3}(a_1+{2})\l(a_1S_{[\sigma}g_{\rho]\lambda}-S_{\sigma}g_{\rho\lambda}\r)+\frac{2}{3}\kappa\l(a_1\Delta_{[\sigma}g_{\rho]\lambda}-\Delta_{\sigma}g_{\rho\lambda}\r)\,.
\eea
We can now split this last expression in its antisymmetric and symmetric part with respect to the two indices $\sigma$ and $\rho$
\bea\label{tofindtorsion}
2(a_3{+1})S_{[\sigma|\lambda|\rho]}+2a_4S_{\sigma\rho\lambda}&=&\Theta_{\lambda\sigma\rho}\,,\nn\\
\label{tofindsymcon}
\partial_{\lambda}g_{\sigma\rho}-\Gamma^{\mu}_{\;\;\rho\lambda}g_{\mu\sigma}-\Gamma^{\nu}_{\;\;\sigma\lambda}g_{\nu\rho}{+2S_{(\sigma|\lambda|\rho)}}&=&\kappa\Delta_{\lambda(\sigma\rho)}-\frac{2}{3}\l[\kappa\Delta_{(\sigma}+(a_1{-1})S_{(\sigma}\r]g_{\rho)\lambda}\nn\\
&&\hspace{-1cm}-g_{\sigma\rho}\Big[\l(\frac{2}{3}a_1-{\frac{2}{3}}\r)S_{\lambda}+\h\kappa \Delta_{\lambda\;\;\mu}^{\;\;\mu}-\frac{1}{3}\kappa\Delta_{\lambda}\Big]\,,\nn\\
\eea
where we have introduced the short hand notation
\be
\Theta_{\lambda\sigma\rho}\equiv\kappa\Delta_{\lambda[\sigma\rho]}+\frac{2}{3}(a_1-1)\l[\kappa\Delta_{[\sigma}+\l(a_1{-1}-\frac{3a_2}{a_1-1}\r)S_{[\sigma}\r]g_{\rho]\lambda}\,.
\ee
Adding suitable permutations of the two expressions in (\ref{tofindsymcon}) we obtain 
\bea\label{torsion}
S_{\rho\nu\mu}\!\!\!&=\!\!\!&\frac{(a_3{+1})}{2(a_3{+1})(a_3+a_4{+1})-4a_4^{\;2}}\l[\Theta_{\nu\rho\mu}-\Theta_{\rho\nu\mu}-\l(2\frac{a_4}{a_3{+1}}-1\r)\Theta_{\mu\rho\nu}\r],\nn\\ \\
\label{gammafin}
\Gamma^{\xi}_{\;\;(\sigma\rho)}\!\!\!&=\!\!\!&\l\{^{\xi}_{\;\;\sigma\rho}\r\}-\h\kappa g^{\xi\lambda}(-\Delta_{\lambda(\sigma\rho)}+\Delta_{\rho(\sigma\lambda)}+\Delta_{\sigma(\rho\lambda)})-\frac{\kappa}{3}\Delta_{(\sigma}\delta_{\rho)}^{\xi}\nn\\
\hspace{-1cm}&&-\frac{\kappa}{4}\l[g_{\sigma\rho}\Delta_{\;\;\mu}^{\xi\;\;\mu}-2\delta^{\xi}_{(\rho}\Delta_{\sigma)\mu}^{\;\;\;\;\;\;\mu}\r]+\frac{2}{3}(a_1-1) S_{(\sigma}\delta^{\xi}_{\rho)}\,.
\eea

Eqs.~(\ref{torsion}) and (\ref{gammafin}) give the antisymmetric and symmetric parts of the connection algebraically in terms of the hypermomentum and the metric. Under the condition that the matter action depends at most linearly on the connection, the above statement is equivalent to saying that we have algebraically expressed the connection in terms of the matter fields and the metric. This assumption is indeed satisfied for all common matter actions, such as scalar and gauge field, in which the matter action does not depend on the connection, and fermions, where the matter action is linear in the connection. This condition can be violated for some fields, for example massive vector fields, especially if not trivial couplings between the connection and the matter are introduced. However, as long as the matter action contains only first order derivatives of the matter fields (in order for the matter fields to satisfy second order equations of motion), $\Delta_{\lambda}^{\;\;\mu\nu}$ will only depend 
algebraically on the connection. This implies that, even though some more complicated manipulations will be required, the connection can always be expressed algebraically in terms of the matter field and the metric (at least at the component level). 

This establishes that the independent connection in (up to) second order metric-affine actions does not carry any dynamics and it can be algebraically eliminated.
Consider now using eqs.~(\ref{gammafin}) and (\ref{torsion}) to completely eliminate the connection in eq.~(\ref{maeq1}). One would then get an equations of the form
\be
\label{grmodsource}
R_{\mu\nu}-\frac{1}{2} R\, g_{\mu\nu}=\kappa \mathcal{T}_{\mu\nu},
\ee
where $R_{\mu\nu}$ and $R$ are the Ricci tensor and the Ricci scalar of the metric $g_{\mu\nu}$ respectively, and $\mathcal{T}_{\mu\nu}$ will be some a second rank tensor which depends on the metric, $\Delta_{\lambda}^{\;\;\mu\nu}$ and $T_{\mu\nu}$. The expression for $\mathcal{T}_{\mu\nu}$ in terms of these three quantities is rather lengthy and we will refrain from writing it here. However, it should already be clear that the theory described by eq.~(\ref{grmodsource}) is general relativity with modified matter interactions. For fields for which the hypermomentum vanishes, $\mathcal{T}_{\mu\nu}=T_{\mu\nu}$.


\section{Higher orders}
\label{higher}

We can now move on to higher orders.  Since the connection has three indices and the derivative one index, there is no $[l^{-3}]$ scalar quantity one can construct out of them. Similarly, one cannot construct an $[l^{-3}]$ scalar quantity using curvature invariants. The next order is $[l^{-4}]$. The terms that could straightforwardly lead to invariants after (several) contractions with the metric are
\bea
\label{list}
&& {\cal R}^{\alpha}_{\phantom{a}\beta\gamma\delta} {\cal R}^{\mu}_{\phantom{a}\nu\lambda\sigma}, \qquad \nabla_\mu \nabla_\nu {\cal R}^{\alpha}_{\phantom{a}\beta\gamma\delta},\qquad {\cal R}^{\alpha}_{\phantom{a}\beta\gamma\delta} S_{\mu\nu}^{\phantom{ab}\lambda} S_{\tau\omega}^{\phantom{ab}\rho},\qquad {\cal R}^{\alpha}_{\phantom{a}\beta\gamma\delta} \nabla_\rho S_{\mu\nu}^{\phantom{ab}\lambda} \nn\\&& S_{\mu\nu}^{\phantom{ab}\lambda} \nabla_\rho {\cal R}^{\alpha}_{\phantom{a}\beta\gamma\delta},\qquad S_{\mu\nu}^{\phantom{ab}\lambda}S_{\alpha\beta}^{\phantom{ab}\sigma}S_{\gamma\delta}^{\phantom{ab}\kappa}S_{\tau\omega}^{\phantom{ab}\rho},  \qquad S_{\mu\nu}^{\phantom{ab}\lambda}S_{\alpha\beta}^{\phantom{ab}\sigma}\nabla_\rho S_{\gamma\delta}^{\phantom{ab}\kappa}\nonumber\\
&&S_{\mu\nu}^{\phantom{ab}\lambda}\nabla_\rho \nabla_\kappa S_{\alpha\beta}^{\phantom{ab}\sigma}, \qquad \nabla_\rho S_{\mu\nu}^{\phantom{ab}\lambda} \nabla_\kappa S_{\alpha\beta}^{\phantom{ab}\sigma}, \qquad \nabla_\mu \nabla_\nu \nabla_\rho S_{\alpha\beta}^{\phantom{ab}\sigma},
\eea
Clearly each of these terms  can lead to various invariants. It goes beyond the purpose of the paper to list all possible terms.\footnote{An exhaustive list of all possible second and fourth order invariants one can construct in the more limiting case where the non-metricity vanishes can be found in \cite{high.ord.terms}. Given that our minimal coupling assumption prevents us from the using of the non-metricity to construct invariants (see also below), this list should cover our case as well.} However, before going further, the following subtle points are worth mentioning:
\begin{enumerate}
\item Due to the symmetries (or lack thereof) of the Riemann tensor when constructed with an independent connection, there are more invariants than in the purely metric case. For example ${\cal R}_{\mu\nu}$ is not symmetric and hence ${\cal R}_{\mu\nu}{\cal R}_{\kappa\lambda}g^{\mu\lambda}g^{\nu\kappa}$ and ${\cal R}_{\mu\nu}{\cal R}_{\kappa\lambda} g^{\mu\kappa}g^{\nu\lambda}$ are not equal.
\item $\nabla_\mu$ is constructed with the independent connection and, hence, total divergences such as $\nabla_\mu u^\mu$ do not lead to pure surface terms and cannot be discarded.
\item Since the metric is not covariantly conserved by the independent connection taking the covariant derivatives first and contracting, or contracting first and then taking a derivative does not lead to the same result. For example the terms $g^{\mu\nu} g^{\alpha\beta} \nabla_\mu \nabla_\nu {\cal R}_{\alpha\beta}$ and $g^{\mu\nu} \nabla_\mu \nabla_\nu {\cal R}$ differ.
\end{enumerate}
Regarding point (ii) one could propose to split the covariant derivative into a metric covariant derivative, which is a surface term, and the rest, such as in (\ref{decomp}). However, writing the rest explicitly would require the use of metric derivatives through the use of the Levi-Civita connection, as discussed above. Something similar can be said about point (iii). The two terms given as an example differ by a term including a covariant derivative of the metric.
This raises the question of whether both of them should be considered. As mentioned earlier, whether terms including derivatives of the metric should be included is really a matter of choice that can be answered only by invoking some minimal coupling principle. If one wants to use the metric purely for contracting indices as suggested previously, then the terms including derivatives of the metric should be suppressed. 

Let us now move on to consider the effect of the higher order terms on the dynamics of the connection. Considering the most general fourth order action is formidable due to the vast number of invariant one would have to include. However, carefully considering isolated terms of different type can still reveal the complete picture.

Clearly there are term in eq.~(\ref{list}) that would not introduce new degrees of freedom if they were added to action (\ref{2ndorder}) as they do not contain extra derivatives, such as the $S^4$ term (indices suppressed). Such terms exist at all even orders, e.g. $S^{2n}$ (again indices suppressed). On the other hand $[l^{-4}]$ terms which contain two derivatives of the Cartan torsion tensor, such as $(\nabla S)^2$ (indices suppressed) would inevitable make the torsion dynamical.

What about fourth order curvature invariants? Let us for the moment set aside the term $\mathcal{R}^2$, since it belongs to the general $f(\mathcal{R})$ class, which we will discuss extensively later, and as we will see it constitutes a rather special case.
A much more characteristic example to consider, which is simple enough to keep calculations tractable and yet general enough to give us the bigger picture is the following 
\be
\label{actionl4}
S=\frac{1}{16\,\pi\,l_p^2} \int dx^4  \sqrt{-g}\left[ {\cal R} +l_p^2 {\cal R}_{\mu\nu}{\cal R}_{\kappa\lambda}(a g^{\mu\kappa}g^{\nu\lambda}+b g^{\mu\lambda}g^{\nu\kappa})\right]+S_M
\ee
As mentioned earlier, when $\mathcal{R}_{\mu\nu}$ is not symmetric, as in our case, the 2 terms in the parenthesis will not lead to the same invariant. In fact, the action can be re-written as
\be
S=\frac{1}{16\,\pi\,l_p^2} \int dx^4  \sqrt{-g}\left[ {\cal R} +l_p^2 c_1 {\cal R}_{(\mu\nu)}{\cal R}^{(\mu\nu)}+ l_p^2 c_2 {\cal R}_{[\mu\nu]}{\cal R}^{[\mu\nu]}\right]+S_M
\ee
where $c_1=a+b$ and $c_2=a-b$.
 This latter form of the action makes the variation easier. The field equations for the metric and the connection are respectively
 \bea
\label{geq}
&&{\cal R}_{(\mu\nu)}-\frac{1}{2} \left({\cal R}+l_p^2 c_1 {\cal R}_{(\alpha\beta)}{\cal R}^{(\alpha\beta)}+ l_p^2 c_2 {\cal R}_{[\alpha\beta]}{\cal R}^{[\alpha\beta]}\right) g_{\mu\nu}\nonumber \\&&\qquad\qquad+2 l_p^2 \,c_1 R_{(\alpha\mu)}R_{(\beta \nu)}g^{\alpha\beta}+2 l_p^2 \,c_2 R_{[\alpha\mu]}R_{[\beta \nu]}g^{\alpha\beta}=\kappa T_{\mu\nu}\,,\\
\label{geq2}
&&\frac{1}{\det}\Big\{-\nabla_\lambda\left[\det g^{\mu\nu}+2\det \left(l_p^2 c_1{\cal R}^{(\mu\nu)}+l_p^2 c_2{\cal R}^{[\mu\nu]}\right) \right]+\nn\\
&&\qquad\qquad+\nabla_\sigma\left[\sqrt{-g}g^{\mu\sigma}+2 \det \left(l_p^2 c_1{\cal R}^{(\mu\sigma)}+l_p^2 c_2{\cal R}^{[\mu\sigma]}\right)\r]\delta_{\;\;\lambda}^{\nu}\Big\}+\nn\\
&&\qquad\qquad+2S^{\;\;\;\;\sigma}_{\lambda\sigma}\l[g^{\mu\nu}+2\left(l_p^2 c_1{\cal R}^{(\mu\nu)}+l_p^2 c_2{\cal R}^{[\mu\nu]}\right)\r]\nn\\&&\qquad\qquad-2S^{\;\;\;\;\sigma}_{\alpha\sigma}\delta^{\nu}_{\;\;\lambda}\l[g^{\mu\alpha}+2\left(l_p^2 c_1{\cal R}^{(\mu\alpha)}+l_p^2 c_2{\cal R}^{[\mu\alpha]}\right)\r]+\nn\\&&\qquad\qquad
+2S_{\alpha\lambda}^{\;\;\;\;\nu} \l[{g^{\mu\alpha}}+2\left(l_p^2 c_1{\cal R}^{(\mu\alpha)}+l_p^2c_2{\cal R}^{[\mu\alpha]}\right)\r]=\kappa\Delta_{\lambda}^{\;\;\mu\nu}\,.
\eea
In the previous section we were able to use the field equation for the connection in order to algebraically express the latter in terms of the metric and the matter fields. Inspecting eq.~(\ref{geq2}), however, one easily realized that, unlike eq.~(\ref{vargamdef}), it appears to include derivatives of the connection due to the presence of $\mathcal{R}_{\mu\nu}$. One could think to use eq.~(\ref{geq}) in order to algebraically express ${\cal R}_{\mu\nu}$ (at least at component level) in terms of the metric and the matter fields (this idea is actually inspired by the specific case of $f(R)$ actions in the more restricted setting of the Palatini formalism where the connection does not couple to the matter --- this  will be discussed below). If this were the case, one could eliminate ${\cal R}_{\mu\nu}$ from  eq.~(\ref{geq2}) and turn it again into an algebraic equation for the connection.

However, this is not possible for generic values of $c_1$ and $c_2$, or $a$ and $b$ for the following simple reasons:
\begin{itemize}
\item ${\cal R}_{\mu\nu}$ is not necessarily symmetric and, therefore, has 16 independent components, whereas eq.~(\ref{geq}) leads to only 10 components equation as it is symmetric in $\mu$ and $\nu$. Therefore, it cannot be used to determine ${\cal R}_{\mu\nu}$ fully, in terms of the metric and the components of $T_{\mu\nu}$.
\item $T_{\mu\nu}$ is not necessarily independent of the connection, as it may include covariant derivatives of certain matter fields. Therefore, even if one would impose such conditions so that eq.~(\ref{geq}) could be solved algebraically to give ${\cal R}_{\mu\nu}$  in term of the metric and  $T_{\mu\nu}$, {e.g.}~impose the constraint ${\cal R}_{[\mu\nu]}=0$ {\em a priori}, that would not actually help in algebraically expressing the connection as a function of the matter fields and the metric (at least for generic matter fields).
\end{itemize}
It should then be clear that the independent connection cannot be eliminated in metric-affine gravity once generic higher order curvature invariants have been added.

The same issue has been considered in Ref.~\cite{STV} for the simpler case of generalized Palatini gravity, {i.e.}~under the assumption that connection does not enter the matter action. This would corresponds to a vanishing $\Delta_{\lambda}^{\;\;\mu\nu}$. The first of the difficulties just discussed is still present in this case when trying to eliminate the connection algebraically by the procedure described above. However, since $T_{\mu\nu}$ is independent of the connection in generalized Palatini gravity, the second difficulty raised here is not really an issue. Hence, it is easier in this framework to write down exceptional Lagrangians for which the connection can be eliminated (it is just an auxiliary field). We refer the reader to Ref.~\cite{STV} for more details. We refrain here from discussing similar exceptions or special cases for metric-affine gravity, as this would require severe fine tuning and/or {\em a priori} constraints. 

Also, we shall not consider explicitly the effect of the mixed terms which include both the Cartan torsion tensor and the Riemann or the Ricci tensor, as this would not add anything new to the qualitative understanding we presented so far. What should be clear by now is that the presence of terms including derivatives of the Cartan torsion tensor or higher order curvature invariants generically leads to a dynamical connection. Therefore,  higher than second order actions generically lead to more dynamical degrees of freedom.


\section{Metric-affine $f(\mathcal{R})$ gravity as a special case}
\label{fofr}

Metric-affine $f({\cal R})$ theories of gravity have been extensively studied lately \cite{T&S}. They constitute a distinct class within higher order actions, in the sense that they allows one to treat terms of different and arbitrarily high order on the same footing. Therefore, even though within the metric-affine setup there is no reason to single out $f({\cal R})$ actions as better motivated ones --- on the contrary, restricting an action to be of this type requires fine tuning --- their simplicity is indeed a good argument for adopting them as toy-models from which to extract general lessons. On the other hand, exactly because they are so special, it is dubious whether $f({\cal R})$ actions can be   considered as representative higher order metric-affine theories from the point of view of their dynamics. This is something that is worth exploring further, which is part of our motivation for considering them separately here.

The other part of our motivation comes from the observation that in the simpler setting of generalized Palatini gravity, where the connection does not enter the matter action,  the whole $f({\cal R})$ class constitutes an exception for which the independent connection does not carry dynamics and can be algebraically eliminated \cite{sot2,Sotiriou:2009xt}. This is true even if the connection is not assumed to be symmetric \cite{Sotiriou:2009xt}. It is, hence, worth exploring in detail what happens in the more general metric-affine framework, in order to avoid confusion and misconceptions.

The action for $f({\cal R})$ theories reads
\be
S=\frac{1}{16\,\pi\,l_p^{\;2}}\int d^4x\det f(\mathcal{R})+S_M
\ee
This action as it stands cannot lead to consistent field equations in the presence of matter, as the gravity part of the action has a symmetry that is not shared by the matter action. The Ricci scalar of the connection ${\cal R}$ remain invariant under the projective transformation
\bea
\Gamma_{\;\;\mu\nu}^{\rho} \rightarrow\Gamma_{\;\;\mu\nu}^{\rho}+\delta_{\;\;\mu}^{\rho}\xi_{\nu}
\eea
($\xi_{\mu}$ being an arbitrary covariant vector field). Consequently any function $f({\cal R})$ and any action of the $f({\cal R})$ type will also be projective invariant. However, matter actions that depend on the connection will not be projective invariant. This has been discussed several times in the literature \cite{hehlkerl,schro, sand,T&S,phd}.

To resolve the inconsistency one needs to somehow break the projective invariance in the gravity sector. The only way to do that, given that we do not want to alter the form of the action, is to constraint the connection to some extent. The meaning of projective invariance is very similar to usual gauge invariance, in the sense that it implies that the connection can be determined only up to a projective transformation. So, to break gauge invariance we need a constraint that that would act as ``gauge fixing''. Clearly, given the nature of the projective transformation we essentially need to fix a vector. It has been argued in Refs.~\cite{T&S,phd} that the best choice for $f({\cal R})$ gravity is  to set 
\bea
S_{\mu}\equiv S_{\alpha\mu}^{\;\;\;\;\alpha}=0
\eea
This constraint can be imposed implicitly, but also explicitly by adding to the action the   Lagrange multiplier 
\bea
S_{LM}=\int d^4 x\det B^{\mu}S_{\mu}\,.
\eea
Varying the total action with respect to metric $g^{\mu\nu}$, connection $\Gamma_{\;\;\mu\nu}^{\rho}$ and Lagrange Multiplier $B^{\mu}$ lead, after some simple manipulations \cite{T&V, T&S},  to the following set of field equations
\be\label{fe1}
f'(\mathcal{R}) \mathcal{R}_{(\mu\nu)}-\frac{1}{2}f(\mathcal{R})g_{\mu\nu}=\kappa T_{\mu\nu}\,,\\
\ee
\bea\label{fe2}
-\nabla_{\lambda}(\sqrt{-g}f'(\mathcal{R})g^{\mu\nu})+\nabla_{\sigma}\l(\sqrt{-g}f'(\mathcal{R})g^{\sigma\mu}\r)\delta^{\nu}_{\;\;\lambda}
+2\sqrt{-g}f'(\mathcal{R})(g^{\mu\nu}S_{\lambda\sigma}^{\;\;\;\;\sigma}\nn\\ 
-g^{\mu\rho}\delta^{\nu}_{\;\;\lambda}S_{\rho\sigma}^{\;\;\;\;\sigma}+g^{\mu\sigma}S_{\sigma\lambda}^{\;\;\;\;\nu})=\kappa\sqrt{-g} \l(\Delta_{\lambda}^{\;\;\mu\nu}-\frac{2}{3}\Delta_{\sigma}^{\;\;\sigma[\nu}\delta^{\mu]}_{\;\;\lambda}\r)\!,
\eea
\be
\label{fe3}
S_{\alpha\mu}^{\;\;\;\;\alpha}=0\,.
\ee
where a prime denotes differentiation with respect to the argument.

{We now check whether it is possible to eliminate algebraically the connection from the field equations. This can be done following a similar procedure as the one used in  section \ref{2order}.} A contraction of eq.~(\ref{fe2}) yields 
\begin{eqnarray}
\nabla_{\sigma}\sqrt{-g}f'(\mathcal{R})g^{\sigma\mu}=\kappa\frac{2}{3}\sqrt{-g}\Delta_{\lambda}^{\;\;(\mu\lambda)}\,.
\end{eqnarray}
We can use this equation in order to eliminate the second term in (\ref{fe2}) to get
\begin{eqnarray}\label{field}
&&-\nabla_{\lambda}(\sqrt{-g}f'(\mathcal{R})g^{\mu\nu})+2\sqrt{-g}f'(\mathcal{R})g^{\mu\sigma}S_{\sigma\lambda}^{\;\;\;\;\nu}=\nonumber \\&&\qquad\qquad=\kappa\sqrt{-g} \l(\Delta_{\lambda}^{\;\;\mu\nu}-\frac{2}{3}\Delta_{\sigma}^{\;\;\sigma[\nu}\delta^{\mu]}_{\;\;\lambda}-\frac{2}{3}\Delta_{\sigma}^{\;\;(\mu\sigma)}\delta^{\nu}_{\;\;\lambda}\r)\,.
\end{eqnarray}
Using the identity
\begin{equation}
g_{\mu\nu}\partial_{\lambda}(\sqrt{-g}f'(\mathcal{R})g^{\mu\nu})=4\sqrt{-g}\partial_{\lambda}f'(\mathcal{R})+2f'(\mathcal{R})\sqrt{-g}\partial_{\lambda}\ln\sqrt{-g}\,,
\end{equation}
and after contracting eq.~(\ref{field}) with the metric in the $\mu$ and $\nu$ indices one gets
\begin{equation}\label{sqrt}
\partial_{\lambda}\ln\sqrt{-g}=\frac{1}{2}\l[-\frac{\kappa}{f'}\l(g_{\mu\nu}\Delta_{\lambda}^{\;\;\mu\nu}-\frac{2}{3}g_{\mu\lambda}\Delta_{\sigma}^{\;\;(\mu\sigma)}\r)-4\frac{\partial_{\lambda}f'}{f'}+2\Gamma^{\sigma}_{\;\;\sigma\lambda}\r]\,.
\end{equation}
Eliminating the density related terms and lowering the indices as we did in the previous section eq.~(\ref{field}) yields
\begin{eqnarray}\label{minchia}
&&-\partial_{\lambda}g_{\alpha\beta}-g_{\alpha\beta}\frac{\partial_{\lambda}f'}{f'}+\Gamma^{\mu}_{\;\;\beta\lambda}g_{\mu\alpha}+\Gamma^{\mu}_{\;\;\lambda\alpha}g_{\mu\beta}=\frac{\kappa}{f'}\Bigg[\frac{1}{2}g_{\alpha\beta}g_{\mu\nu}\Delta_{\lambda}^{\;\;\mu\nu}\\&&\qquad-\frac{1}{3}g_{\alpha\beta}g_{\mu\lambda}\Delta_{\sigma}^{\;\;(\mu\sigma)}-g_{\mu\alpha}g_{\nu\beta}\Delta_{\lambda}^{\;\;\mu\nu}+\frac{1}{3}\l(g_{\alpha\lambda}g_{\nu\beta}\Delta_{\sigma}^{\;\;\sigma\nu}+g_{\lambda\beta}g_{\mu\alpha}\Delta_{\sigma}^{\;\;\mu\sigma}\r)\Bigg]\,.\nn
\end{eqnarray}
Adding suitable permutations of eq.~(\ref{minchia}) one gets
\begin{eqnarray}\label{gamma}
\Gamma^{\rho}_{\;\;\alpha\beta}=\l\{^\rho_{\phantom{a}\alpha\beta}\r\}+\frac{1}{2f'}\l[\partial_{\alpha}f'\delta^{\rho}_{\;\;\beta}+\partial_{\beta}f'\delta^{\rho}_{\;\;\alpha}-g^{\rho\lambda}g_{\alpha\beta}\partial_{\lambda}f'\r]+\frac{\kappa}{f'}W_{\alpha\beta}^{\;\;\;\;\rho}\,,
\end{eqnarray}
where $\l\{^\rho_{\phantom{a}\alpha\beta}\r\}$ is the usual Levi-Civita connection associated with the metric $g_{\mu\nu}$ and $W_{\alpha\beta}^{\;\;\;\;\rho}$ is a tensor encompassing all the hypermomenta terms
\bea
W_{\alpha\beta}^{\;\;\;\;\rho}=&&-\frac{1}{2}\l\{\frac{1}{2}g_{\alpha\beta}g_{\mu\nu}\Delta^{\rho\mu\nu}-g_{\mu\nu}\delta^{\rho}_{\;\;(\alpha}\Delta_{\beta)}^{\;\;\mu\nu}+\Delta_{\beta\;\;\alpha}^{\;\;\rho}+\Delta_{\alpha\beta}^{\;\;\;\;\rho}\r.\\&&-\Delta^{\rho}_{\;\;\alpha\beta}-g_{\alpha\beta}\Delta_{\sigma}^{\;\;(\rho\sigma)}+\frac{1}{3}\delta^{\rho}_{\;\;\alpha}g_{\mu\beta}\l(2\Delta_{\sigma}^{\;\;[\sigma\mu]}+\Delta_{\sigma}^{\;\;(\sigma\mu)}\r)\nonumber\\&&\l.+\frac{1}{3}\delta^{\rho}_{\;\;\beta}g_{\mu\alpha}\l(2\Delta_{\sigma}^{\;\;[\sigma\mu]}+\Delta_{\sigma}^{\;\;(\sigma\mu)}\r)\r\}\,.\nn
\eea

Eq.~(\ref{gamma}) provides and expression for the connection in terms of the metric, the hypermomentum but also ${\cal R}$, via the presence of $f$. So, we essentially run into the same difficulties we faced in the previous section when trying to eliminate the connection. However, here ${\cal R}$ is just a scalar quantity.
Consider the trace of eq.~(\ref{fe1})
\be\label{tracefe1}
\mathcal{R}f'(\mathcal{R})-2f(\mathcal{R})=\kappa T\,.
\ee
For a given function $f$ this is an algebraic equation in ${\cal R}$. 
Setting aside pathological cases in which this equation has no root, and the exceptional case where  $f(\mathcal{R})\propto\mathcal{R}^2$, which corresponds to a conformally invariant gravitational action (see Refs.~\cite{fer,sot1,T&S} for more details), eq.~(\ref{tracefe1}) can be used to express $\mathcal{R}$ as an algebraic function of $T$. This expression can in turn be used to eliminate ${\cal R}$ in favour of $T$ in the $f$ terms in eq.~(\ref{gamma}). Therefore, from now on we can be thinking of eq.~(\ref{gamma}) as expressing the affine connection as a function of just derivatives of metric, $T_{\mu\nu}$ and the hypermomentum. This mean we are clear of the first difficulty encountered for generic fourth order actions.

This is not the case for the second point we made previously though, {\em i.e.}~that $T_{\mu\nu}$, and hence $T$, can generically depend on the connection. Even though the requirement that the matter satisfies second order differential equations of motion essentially implies that the dependence of $T_{\mu\nu}$ on the connection will be algebraic (there can be only first covariant derivatives of the matter fields is $S_M$), the fact that there are first order derivatives of $f'$ in eq.~(\ref{gamma}) is enough to give derivatives of the connection. Therefore, in metric-affine $f({\cal R})$ the connection satisfies a dynamical equation in general.

A remarkable observation is the following. Taking the antisymmetric part of (\ref{gamma}) in its lower indices  we get
\bea
\Gamma^{\rho}_{\;\;[\alpha\beta]}\equiv S_{\alpha\beta}^{\phantom{ab}\rho}&=&\Delta_{[\beta\;\;\alpha]}^{\;\;\;\rho}+\Delta_{[\alpha\beta]}^{\;\;\;\;\;\;\rho}-\Delta^{\rho}_{\;\;[\alpha\beta]}\\&=&g^{\rho\lambda}\l(\Delta_{\beta[\lambda\alpha]}+\Delta_{\alpha[\beta\lambda]}-\Delta_{\lambda[\alpha\beta]}\r)\,.\nn
\eea
This implies that the torsion is still non-dynamical and vanishes for matter fields with vanishing $\Delta_{\gamma}^{\;\;[\alpha\beta]}$. It is only the symmetric part of the connection that carries dynamics. As already stressed in \cite{T&S} torsion is non-propagating in metric-affine $f({\cal R})$ and it is introduced by matter fields having $\Delta_{\gamma}^{\;\;[\alpha\beta]}\neq0$.

The fact that the connection appears to satisfy a first order differential equation, at least if one assumes that $T$ does not include any derivatives of the connection, seems worrying. However, it is very difficult to tell if this is indeed a problem. Neither do we have the exact form of the equation, nor do we know which degrees of freedom hiding in the connection will actually be excited.

Of course, for matter fields which do not couple to the connection (scalar field, gauge fields) or if one imposes that the independent connection does not enter the matter action $S_M$, $T_{\mu\nu}$ is independent of the connection as well and $\Delta_{\lambda}^{\;\;\mu\nu}=0$. In this case the connection can indeed be eliminated and one recovers the results of Palatini $f({\cal R})$ gravity \cite{Sotiriou:2009xt}. Another special case is the one where $f({\cal R})={\cal R}$, as is this case $f'=1$ and ${\cal R}$ is no longer present in the eq.~(\ref{gamma}), which now takes the form
\begin{eqnarray}\label{gammaR}
\Gamma^{\rho}_{\;\;\alpha\beta}=\l\{^\rho_{\phantom{a}\alpha\beta}\r\}+\kappa\,W_{\alpha\beta}^{\;\;\;\;\rho}\,.
\end{eqnarray}
One can then write
\bea
\label{ricciR}
\hspace{-0.5cm}\cal{R}_{(\alpha\beta)}&\equiv&\partial_{\rho}\Gamma_{\;\;(\alpha\beta)}^{\rho}-\partial_{(\beta}\Gamma_{\;\;\alpha)\rho}^{\rho}+\Gamma_{\;\;\sigma\rho}^{\rho}\Gamma_{\;\;(\alpha\beta)}^{\sigma}-\Gamma_{\;\;\sigma(\beta}^{\rho}\Gamma_{\;\;\alpha)\rho}^{\sigma}=\\
\hspace{-0.5cm}&=&R_{\alpha\beta}+\kappa\left[\bar{\nabla}_\rho W_{(\alpha\beta)}^{\;\;\;\;\;\;\rho}-\bar{\nabla}_{(\beta}W_{\alpha)\rho}^{\;\;\;\;\;\;\rho}+W_{\sigma\rho}^{\;\;\;\;\;\;\rho}W_{(\alpha\beta)}^{\;\;\;\;\;\;\sigma}-W_{\sigma(\beta}^{\;\;\;\;\;\;\rho}W_{\alpha)\rho}^{\;\;\;\;\;\;\sigma}\right]\,,\nn
\eea
where $R_{\mu\nu}$ is the Ricci tensor of the metric $g_{\mu\nu}$ and $\bar{\nabla}_\mu$ is the covariant derivative defined with the Levi-Civita connection of the same metric. Contracting with the metric one gets
\bea
\label{rsR}
{\cal R}&=&R+\kappa\left[2 \bar{\nabla}_{[\rho}W_{\phantom{a}\mu]}^{\mu\phantom{a}\rho}+W_{\sigma\rho}^{\phantom{ab}\rho}W_{\mu}^{\phantom{a}\mu\sigma}-W_{\sigma}^{\phantom{a}\mu\rho}W_{\mu\rho}^{\phantom{ab}\sigma}\right]\,.
\eea
We can now use eqs.~(\ref{gammaR}), (\ref{ricciR}) and (\ref{rsR}) in order to completely eliminate the connection and end up with the single field equation for the metric
\bea
\label{eq:field}
G_{\alpha\beta}&&=\kappa\, T_{\alpha\beta}+\frac{\kappa}{2}g_{\alpha\beta}\l\{2 \bar{\nabla}_{[\rho}W_{\;\;\mu]}^{\mu\;\;\;\;\rho}+W_{\sigma\rho}^{\;\;\;\;\;\;\rho}W_{\mu}^{\;\;\mu\sigma}-W_{\sigma}^{\;\;\mu\rho}W_{\mu\rho}^{\;\;\;\;\;\;\sigma}\r\}\nonumber\\
&&-\kappa\l\{\bar{\nabla}_\rho W_{(\alpha\beta)}^{\;\;\;\;\;\;\rho}-\bar{\nabla}_{(\beta}W_{\alpha)\rho}^{\;\;\;\;\;\;\rho}+W_{\sigma\rho}^{\;\;\;\;\;\;\rho}W_{(\alpha\beta)}^{\;\;\;\;\;\;\sigma}-W_{\sigma(\beta}^{\;\;\;\;\;\;\rho}W_{\alpha)\rho}^{\;\;\;\;\;\;\sigma}\r\},
\eea
where, as usual,
\be
G_{\alpha\beta}\equiv R_{\alpha\beta}-\frac{1}{2} R\, g_{\alpha\beta}\,,
\ee
is the Einstein tensor of the metric $g_{\mu\nu}$. Therefore, $f({\cal R})={\cal R}$ metric-affine gravity reduces to general relativity with extra matter interactions. This is anyway clearly just a subcase of the most general second order action we examined in section \ref{2order} with vanishing $a_i$'s.

However, we have shown for any other function $f({\cal R})$ the connection cannot be algebraically eliminated in the presence of matter fields that couple to it.


\section{Discussion and Conclusions}
\label{discuss}

We have studied the dynamics of theories of gravity in which the metric and the connection are independent quantities. Instead of restricting ourselves to a specific action, which would inevitably affect the generality of our conclusions, we chose to follow an approach inspired by effective field theory and attempt to understand how are the dynamics of the theory affected when increasing the order of the various invariant included in the action. To this end we first considered the most general action formed by second order invariants and then moved on to examine how these would be modified by including different types of higher order terms in the action. In both cases we imposed a generalized minimal coupling principle in order to reduce the number of terms to be considered, which excludes invariants constructed with the non-metricity or the metric curvature.

Our main conclusions are the following:
\begin{itemize}
\item Even for the most general action one can construct with second order invariants the connection does not carry any dynamics and can always be algebraically eliminated. That is, at this order, metric-affine gravity can always be written as general relativity with a modified source term or extra matter interactions. No extra degrees of freedom are excited.
\item Including higher order terms in the action changes the situation radically. The connection (or parts of it)  becomes dynamical and so, it cannot be eliminated algebraically. The theory now propagates more degrees of freedom than general relativity. Thus, seen as an effective field theory, metric-affine gravity is rather peculiar and its dynamics can deceive: at the lowest order the extra degrees of freedom appear to lose their dynamics and become auxiliary fields, but once higher order terms are taken into account the extra degrees of freedom do propagate. To avoid exciting extra degrees of freedom significant fine tuning and extra {\em a priori} constraints are required.
\item $f(\mathcal{R})$ actions, which have been previously considered in metric-affine gravity, appear to constitute a distinct class with special properties. Even though the connection does carry dynamics in the presence of fields coupling to it --- unlike the simplified case of Palatini $f({\cal R})$ gravity --- torsion remains non-propagating. The propagating degrees of freedom reside only in the symmetric part of the connection.  In this sense, $f({\cal R})$ actions cannot be considered representative examples of generic higher order metric-affine theories.
\end{itemize}

From an effective field theory perspective it seems that there are dynamical degrees of freedom in metric-affine gravity which appear to ``freeze" at low energies and can be eliminated in favour of extra matter interaction. This implies that a possible low energy manifestation of metric-affine gravity could be revealed in matter experiments in terms of such interactions, but the phenomenology of metric-affine theories is not limited to that. It is much richer and it includes extra propagating degrees of freedom, which can potentially be detected. A typical, but certainly not the only, example would be the presence of propagating torsion, whose consequences have been studied in a limiting setting \cite{carroll}.\footnote{See also Ref.~\cite{Shapiro:2010zq}, which appeared during the completion of this manuscript, and references therein.} 

It goes beyond the scope of the current study to examine further the phenomenology of metric-affine gravity. It would be very interesting to understand in more detail how the extra degrees of freedom behave in the regime where they are dynamical and how exactly do they modify matter interactions at low energies. It is also crucial to examine the predictions of such theories for energy conservation and violations of the various formulations of the equivalence principle. Such considerations would allow us to place constraints on metric-affine theories. 

 A probable point of concern can be our use of the generalized minimal coupling principle. One could argue that it is not compatible with our effective field theory perspective as radiative corrections would not respect such a principle. One could also feel uneasy treating non-metricity and torsion on a different footing. Indeed, the minimal coupling principle is used here mostly as a way to reduce the number of terms one has to take into consideration and it should not necessarily be considered as a fundamental principle. Abandoning it and considering the most general action possible is the next step.

As a closing remark, we would like to mention the following. Clearly, one might question how fundamental is the geometrical interpretation of metric-affine gravity. In fact, since for second order actions one can always eliminate the independent connection, the latter can be regarded as an auxiliary field. Even for actions with higher order terms though, where degrees of freedom residing in the connection will be excited, one could have an equivalent representation without an independent affine connection (recall that an independent connection can always be written as the Levi-Civita connection plus a tensor). Indeed, which representation one choose is a matter of preference, at least at a classical level, as the dynamical content of the theory is one and the same. On the other hand, it is worth pointing out that the choice of representations becomes a factor when constructing the action of the theory. It influences our choices regarding the presence of some terms by making some exclusion principles, such as 
minimal coupling and its generalizations, more or less appealing (see also Ref.~\cite{Sotiriou:2007zu} for a more general discussion on this issue). This is a subtle point that needs to be taken seriously into account.


\section*{Acknowledgments}

The authors would like to thank Roberto Percacci for drawing their attention to a field approach towards metric-affine gravity and Yuri Obukhov for discussions. TPS was supported in part by the STFC and in part by a Marie Curie Incoming International Fellowship.
\\\\
\noindent {\bf Note added:} In the published version of this manuscript a term was omitted in the variation of actions (14) and (34). This omission had led to missing terms in equations (16), (19)-(22), (25)-(30) and (36), which were later corrected in an erratum. These corrections have been directly incorporated in this preprint.

\end{document}